\documentclass{eas}
\usepackage{graphicx}
\begin{document}

\title{Status of the EDELWEISS-II experiment} 
\author{V. SANGLARD on behalf of the EDELWEISS Collaboration}\address{Universit\'e de Lyon, F-69622, Lyon, France;
Universit\'e Lyon 1, Villeurbanne;
CNRS/IN2P3, UMR5822, Institut de Physique Nucl\'eaire de Lyon}
\begin{abstract}
EDELWEISS is a direct dark matter search experiment situated in the low
radioactivity environment of the Modane Underground Laboratory.
The experiment uses Ge detectors at very low temperature in order to
identify eventual rare nuclear recoils induced by elastic scattering of
WIMPs from our Galactic halo.
We present results of the commissioning of the second phase of the
experiment, involving more than 7 kg of Ge, that has been completed in
2007. 
We describe two new types of detectors with active rejection of events
due to surface contamination. This active rejection is required in order to achieve the physics goals of 10$^{-8}$ pb cross-section measurement for the current phase. 
\end{abstract}
\maketitle
\section{Introduction}
Recent cosmological observations of the CMB show that the main part of the 
matter in our Universe is dark and non baryonic (Hinshaw {\em et al.}  \cite{WMAP}). If non baryonic Dark Matter is made of particles,
they must be stable, neutral and massive : WIMPs (Weakly Interactive Massive Particles). 
In the MSSM (Minimal Supersymmetric Standard Model) framework, the WIMP could be the LSP 
(Lightest Supersymmetric Particle) called neutralino. It has a mass between few tens and 
few hundreds of GeV/c$^2$, and a scattering cross-section with a nucleon below $10^{-6}$~pb.\\  
The EDELWEISS (Exp\'erience pour D\'etecter les WIMPs en Site Souterrain) experiment is dedicated to the direct detection of WIMPs. 
The direct detection principle (used also by other experiments like CDMS (Akerib {\em et al.} \cite{cdms}), CRESST (Angloher {\em et al.}  \cite{cresst}) and XENON (Angle {\em et al.}  \cite{xenon}))
consists in the measurement of the energy released by nuclear recoils produced in an ordinary matter target 
by the elastic collision of a WIMP from the galactic halo.\\
The main challenge is the expected extremely low event rate ($\leq$ 1 evt/kg/year) due to the very small interaction cross-section of WIMP with  nucleons. An other constraint is the relatively small deposited energy  ($\leq$ 100 keV). 
\section{Experimental set-up and detectors}
\begin{figure}[htb]
\begin{center}
\includegraphics*[width=9cm,height=6cm]{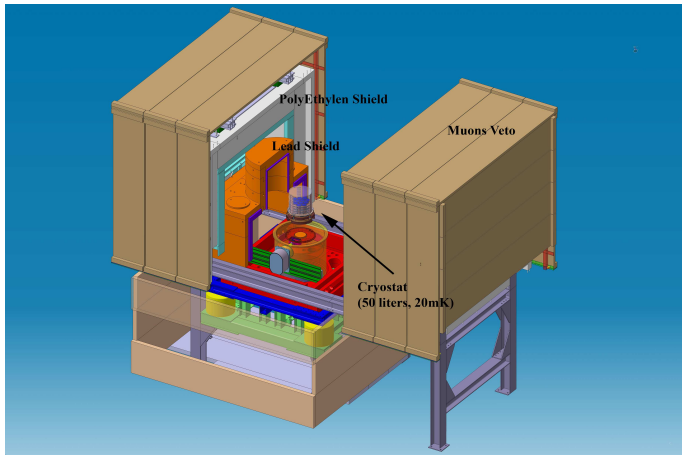}
\caption{General scheme of the EDELWEISS-II experiment. The outer shell is the muon veto system, followed by the neutron polyethylen shield and the inner lead shield. The upper part can be open to access the bolometers, while the cryogenic systems are located under the detectors.}
\label{fig:setup}
\end{center}
\end{figure}

The EDELWEISS experiment is located under $\sim$1700~m of rock ($\sim$4800~mwe) in the Modane Underground Laboratory (LSM) in the highway Fr\'ejus tunnel 
connecting France and Italy 
In the laboratory, the muon flux is reduced down to 4~$\mu$/m$^2$/d, a factor $10^6$ times less than at the surface.\\
The EDELWEISS-II experiment installation (see Fig.~\ref{fig:setup}) was completed end of 2005. Specific improvements have been made in order to reduce the possible background sources that have limited the sensitivity of the previous experiment EDELWEISS-I (Fiorucci  {\em et al.} \cite{bkgd} ; Sanglard  {\em et al.} \cite{final} ).
To reduce the radioactive background in the cryostat all the materials were tested for 
radiopurity in a HPGe dedicated detector with very low radon level. The experiment is installed in a class 10 000 clean room  
and the cryostat environment is submitted to a permanent flow of deradonized air. 
The gamma background is screened by a 20 cm thick lead shield.
Concerning the low energy neutron background, due to the radioactive surrounding rock, it is attenuated 
by more than three orders of magnitude thanks to a 50 cm polyethylene shield. 
In addition, a muon veto wraps the experiment though muons interacting
in the lead shield are tagged. \\
The dilution cryostat is of inverted design, with the experimental detector volume on the top. The large volume, 50 $\ell$, allows for the installation up to 120 identical detectors in a compact arrangement, that will improve the possibility of 
detecting multiple interactions of neutrons and hence reject them. Simulations show a nuclear recoil rate above 10 keV  estimated to be  less than 10$^{-3}$ evt/kg/d, corresponding to a WIMP-nucleon cross-section sensitivity of $10^{-8}$ pb for a WIMP mass of $\sim$ 100 GeV/c$^2$ corresponding to an improvement of a factor 100 compared to EDELWEISS-I. \\
The detectors used in the experiment are high purity germanium crystals with measurement of
phonon and ionization signals, cooled at a temperature of $\sim$ 20 mK. The simultaneous measurement of both heat and ionization signals provides an excellent 
event by event discrimination between nuclear recoils (induced by WIMP or neutron 
scattering) and electron recoils  (induced by $\alpha$, $\beta$ or $\gamma$-radioactivity). The ratio
of the ionization to the heat signals depends on the recoiling particle, since a nucleus
produces less ionization than an electron does.\\
One important limitation for the EDELWEISS-I sensitivity was the presence of surface events, namely interactions occurring near electrodes. 
Because of diffusion, trapping and recombination, the charge induced by surface events is miscollected and can mimic nuclear recoils. To reach expected WIMP-nucleon cross-sections, it is necessary to have an active rejection in identifying the surface events. 
For this purpose, the experiment is running with three types of detectors : classic EDELWEISS-I design 320 g Ge/NTD equipped with new Teflon holders, 400 g Ge/NbSi equipped with two NbSi sensors sensitive to athermal phonons (Juillard {\em et al.}  \cite{nbsi1} ; Marnieros {\em et al.}  \cite{nbsi2}) and 400 g Ge/NTD/ID crystal equipped with interdigitized electrode scheme (Broniatowski {\em et al.}  \cite{id1} ; Defay {\em et al.}  \cite{id2}) (described in the next section).\\ 
Also, an integration test is underway of a scintillation-phonon detector in EDEL\-WEISS-II (Di Stefano {\em et al.}  \cite{ias}).  Such detectors could offer additional target materials, and assist understanding of backgrounds and demonstration of a signal.

\section{Commissioning runs}
Comparing to EDELWEISS-I, the EDELWEISS-II experiment is completely new with new cryostat, new electronics, new acquisition hardware and software. Consequently the year 2006 has been dedicated to the tuning of electronics, acquisition and cryogenics with 8 detectors. In 2007, commissioning runs were performed with 25 detectors (21 Ge/NTD and 4 Ge/NbSi) with gamma and neutron calibrations. Low background runs were also performed in order to study the $\alpha$, $\beta$ and $\gamma$ backgrounds.  \\

\begin{figure}[htb]
\begin{center}
\includegraphics*[width=7cm,height=6cm]{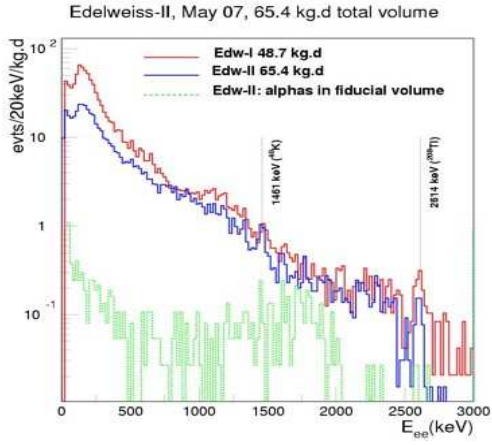}
\caption{Ionization energy spectrum for low background runs measured by Ge/NTD detectors (65 kg.d for the total volume). The alpha background is identified by selecting events with ionization/recoil energy ratio less than 0.5.}
\label{fig:fondgamma}
\end{center}
\end{figure}
Fig.~\ref{fig:fondgamma} shows the electronic recoil background rate in the fiducial volume (Martineau {\em et al.}  \cite{volfid}). For energies below 100 keV, the rate is approximately 0.6 evt/keV/kg/d which is a factor 2.5 better than EDELWEISS-I. The alpha rate background is between 1.6 and 4.4 $\alpha$/kg/d, depending on the detector and its near-environment (for EDELWEISS-I the mean alpha rate was 4.2 $\alpha$/kg/d).\\

\begin{figure}[htb]
\begin{center}
\includegraphics*[width=10cm,height=6cm]{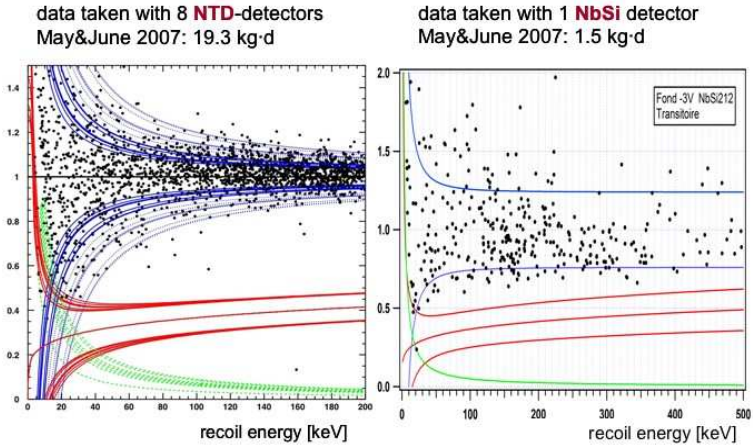}
\caption{Distribution of the ionization/recoil energy ratio as a function of recoil energy for low background runs. Left part : for 8 Ge/NTD detectors with a fiducial exposure of 19.3 kg.d - Right part : for 1 Ge/NbSi detector with a fiducial exposure of 1.5 kg.d.}
\label{fig:lowbckgd}
\end{center}
\end{figure}
Fig.~\ref{fig:lowbckgd} shows the ionization/recoil energy ratio as a function of the recoil energy for low background runs taken at the end of the commissioning runs. On the left, results for the 8 Ge/NTD detectors with lowest threshold (between 20 and 35 keV) are shown for a fiducial exposure of 19.3 kg.d. No events have been observed in the nuclear recoil band suggesting a possible improvement compared to EDELWEISS-I. But more statistics are needed to draw firm conclusions. A fiducial exposure of $\sim$ 100 kg.d obtained in stable conditions is currently being analyzed.\\ 
On the right part of Fig.~\ref{fig:lowbckgd} are shown results for a 200 g Ge/NbSi detector corresponding to a fiducial exposure of 1.5 kg.d after cuts removing surface events. 
On this figure, the recoil energy is estimated using the fast athermal signal, which gives the best resolution at low energy in this detector, and is also used to detect near-surface events 
(Marnieros {\em et al.}  \cite{nbsi2}). However the dispersion at high energy is too large and further improvements are needed before using this type of detector for WIMP searches.\\
\begin{figure}[htb]
\begin{center}
\includegraphics*[width=7cm,height=7cm]{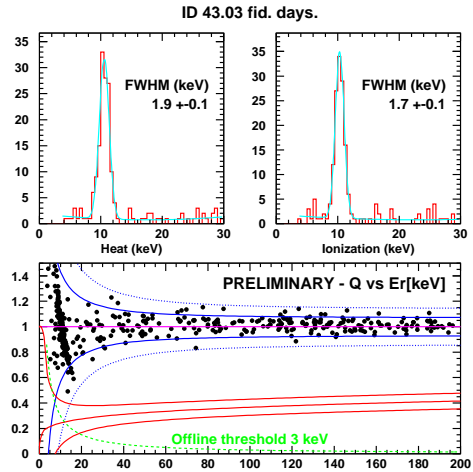}
\caption{Results obtained with a 200 g Ge/NTD/ID detector at LSM - Top : Heat and ionization resolutions - Bottom : Distribution of the ionization/recoil energy ratio as a function of recoil energy for low background runs with a fiducial exposure of few kg.d.}
\label{fig:id}
\end{center}
\end{figure}

In november 2007, a new type of detector has been installed in the EDELWEISS-II cryostat for which the standard ionization electrodes are replaced by interdigitized (ID) electrodes with specific voltage modifying the topology of the electric field near surface (Broniatowski {\em et al.}  \cite{id1}). The signals from the so-called "field-shaping" electrodes located on both surfaces of the detector provide an active rejection of surface events. Preliminary test of a 200 g detector at ground level showed a surface event rejection better than 99.5 $\%$. 
Fig.~\ref{fig:id} shows the data recorded in the EDELWEISS-II set-up with a fiducial exposure of 43 days. Resolutions are below 2 keV for both heat and ionization channels. No events are observed in the nuclear recoil band down to a threshold of $\sim$ 15 keV.

 \section{Conclusions}
The EDELWEISS-II set-up has been validated with calibration and low background runs. Energy resolutions and discrimination capabilities close to EDEL\-WEISS-I results have been measured for Ge/NTD detectors. Analysis of $\sim$ 100 kg.d of fiducial exposure are underway with the objective of quantifying the improvement of the identification of low energy surface event contamination of the detectors.

In parallel to the operation of classic EDELWEISS-I type Ge/NTD detectors, two types of detectors with active rejection of surface events have been tested. While progress are still needed for Ge/NbSi detectors, considerable advances have been made in the validation of the Ge/NTD/ID detectors, both in surface laboratory and at LSM. Since may 2008, the 200 g Ge/NTD/ID detector has been equipped with a $^{210}$Pb source to measure the surface event rejection and three other 400 g Ge/NTD/ID detectors have been installed in the EDELWEISS-II cryostat. \\
Thank to detectors with active surface event rejection and the increase fiducial mass, a sensitivity of few 10$^{-9}$ pb could be reached in the coming years.   



\begin{thebibliography}{99}
\bibitem[2009]{WMAP} G. Hinshawl et al., accepted in Astrophys. J. Suppl. S. (2009), astro-ph/0803.0732
\bibitem[2006]{cdms} D.S. Akerib et al., {\it Phys. Rev. Lett}, {\bf 96}, 011302 (2006)
\bibitem[2005]{cresst} G. Angloher et al., {\it Astropart. Phys.}, {\bf 24}, 325 (2005) 
\bibitem[2008]{xenon} J. angle et al., {\it Phys. Rev. Lett}, {\bf 101}, 091301 (2008) 
\bibitem[2007]{bkgd} S. Fiorucci et al., {\it Astropart. Phys.}, {\bf 28}, 143 (2007) 
\bibitem[2005]{final} V. Sanglard et al., {\it Phys. Rev. D}, {\bf 71}, 122002 (2005) 
\bibitem[2006]{nbsi1} A. Juillard et al., {\it Nucl. Instrum. Methods Phys. Res. Sec. A}, {\bf 559}, 393 (2006) 
\bibitem[2008]{nbsi2} S. Marnieros et al., {\it J. Low Temp. Phys.}, {\bf 151}, 835 (2008) 
\bibitem[2008]{id1} A. Broniatowski et al., {\it J. Low Temp. Phys.}, {\bf 151}, 830 (2008) 
\bibitem[2008]{id2} X. Defay et al., {\it J. Low Temp. Phys.}, {\bf 151}, 896 (2008)
\bibitem[2008]{ias} PCF Di Stefano et al., {\it J. Low Temp. Phys.}, {\bf 151}, 902 (2008) 
\bibitem[2004]{volfid} O. Martineau et al., {\it Nucl. Instrum. Methods Phys. Res. Sec. A}, {\bf 530}, 426 (2004) 
\end{thebibliography}
\end{document}